# Variational Formulation of the Log-Aesthetic Surface and Development of Discrete Surface Filters

Kenjiro T. Miura[1], Ryo Shirahata[2], Shin'ichi Agari[3], Shin Usuki[4] and R.U. Gobithaasan[5]

[1]Shizuoka University, tmkmiur@ipc.shizuoka.ac.jp
[2]Shizuoka University, f0930018@ipc.shizuoka.ac.jp
[3]Shizuoka University, f5945014@ipc.shizuoka.ac.jp
[4]Shizuoka University, dsusuki@ipc.shizuoka.ac.jp
[5]University Malaysia Terengganu, gr@umt.edu.my

## ABSTRACT

The log-aesthetic curves include the logarithmic (equiangular) spiral, clothoid, and involute curves. Although most of them are expressed only by an integral form of the tangent vector, it is possible to interactively generate and deform them and they are expected to be utilized for practical use of industrial and graphical design. The discrete log-aesthetic filter based on the formulation of the log-aesthetic curve has successfully been introduced not to impose strong constraints on the designer's activity, to let him/her design freely and to embed the properties of the log-aesthetic curves for complicated ones with both increasing and decreasing curvature. In this paper, in order to define the log-aesthetic surface and develop surface filters based on its formulation, at first we reformulate the log-aesthetic curve with variational principle. Then we propose several new functionals to be minimized for free-form surfaces and define the log-aesthetic surface. Furthermore we propose new discrete surface filters based on the log-aesthetic surface formulation.

## 1 INTRODUCTION

Thanks to the developments of the measurement and information technology, the reverse engineering is utilized which generate digital models on the computers from 3 dimensional physical models in clay or wood. Because free-form surfaces are frequently used for aesthetic design and their measurement data are usually huge and they have errors, it is very difficult to generate a high-quality digital model with smooth changes of curvature from such data.

The log-aesthetic curves include the logarithmic (equiangular) curve (the slope of the LCG: logarithmic curvature graph $\alpha = 1$ ), the clothoid curve ($\alpha = -1$), the circle involute ($\alpha = 2$ ) and Nielsen's spiral ($\alpha = 0$). Recently the generalized Cornu spiral [12] has been reported to include several log-aesthetic curves since its curvature profile is given by a rational linear function and so its LCG gradient is given by a straight line function [11]. It is possible to generate and deform the log-aesthetic curve in real time even if they are expressed by integral forms using their unit tangent vectors as integrands ($\alpha \neq 1, 2$ ) and they are expected to be used in practical applications [1, 20]. The discrete log-aesthetic filter based on the formulation of the log-aesthetic curve has successfully been introduced not to impose strong constraints on the designer's activity, to let him/her design freely and to embed the properties of the log-aesthetic curves for complicated curves with both increasing and decreasing curvature [21].

Therefore, in this paper at first we define the log-aesthetic surface in two ways, one utilizes the self-affinity of the surface and the other is based on variational principle. Then we propose a discrete filter named log-aesthetic surface filter that removes noises from a set of points obtained by a 3D laser range scanner, smooth them out and make the surface log-aesthetic as well.

The rest of the paper is organized as follows. Section 2 describes related work and sections 3 discusses about the self-affinity of the surface and the formulation of the log-aesthetic surface using it. Section 4 explains the formulation of the log-aesthetic surface based on variational principle. Section 5 introduces a new discrete surface filters based on the log-aesthetic surface. Finally, we conclude the paper in section 6 with a discussion of future work.

## 2  PRELATED WORK

In this section, we discuss related researches on the log-aesthetic curve, curvature based energy functionals for fair surfaces, and discrete filters.

### 2.1  Log-aesthetic Curve

 "Aesthetic curves" were proposed by Harada et al. [7] as such curves whose logarithmic distribution diagram of curvature (LDDC) is approximated by a straight line. Miura et al. [17, 18] derived analytical solutions of the curves whose logarithmic curvature graph (LCG): an analytical version of the LDDC [7] are strictly given by a straight line and proposed these lines as general equations of aesthetic curves. Furthermore, Yoshida and Saito [26] analyzed the properties of the curves expressed by the general equations and developed a new method to interactively generate a curve by specifying two end points and the tangent vectors there with three control points as well as : the slope of the straight line of the LCG. In this research, we call the curves expressed by the general equations of aesthetic curves the log-aesthetic curves.

The problems of the connection of plural log-aesthetic segments was dealt by Miura et al. [20] and an input method of the compound-rhythm log-aesthetic curve which consist of two log-aesthetic curve segments connected with $C^3$ continuity was proposed by Agari [1]. Furthermore an extension of the planar log-aesthetic curve into space: the log-aesthetic space curve was proposed by Miura et al. [19] and it was classified by Yoshida and Saito [27]. This section discusses several important properties of log-aesthetic curves. Note that an aesthetic curve is a curve whose logarithmic curvature graph is given by a straight line.

#### 2.1.1  General equations of aesthetic curves

For a given curve, we assume the arc length of the curve and the radius of curvature are denoted by $s$ and $\rho$, respectively. The horizontal axis of the logarithmic curvature graph measures $\boldsymbol{log}\,\rho$ and the vertical axis measures $\boldsymbol{log}\big(ds/d(\boldsymbol{log}\,\rho)\big)=\boldsymbol{log}\big(\rho\,ds/d\rho\big)$. If the LCG is given by a straight line, there exists a constant $\alpha$ such that the following equation is satisfied:

$$\boldsymbol{log}\left(\rho\frac{ds}{d\rho}\right)=\alpha\,\boldsymbol{log}\,\rho+C \tag{2.1}$$

where $C$ is a constant. The above equation is called the fundamental equation of aesthetic curves [8]. Rewriting Eqn. (2.5), we obtain:

$$\frac{1}{\rho^{\alpha-1}}\frac{ds}{d\rho}=e^{C}=C_{0} \tag{2.2}$$

Hence there is some constant $c_0$ such that:

$$\rho^{\alpha-1}\frac{d\rho}{ds}=c_{0} \tag{2.3}$$

From the above equation, when $\alpha\neq 0$, the first general equation of aesthetic curves

$$\rho^{\alpha}=c_{0}s+c_{1} \tag{2.4}$$

is obtained. If $\alpha=0$, we obtain the second general equation of aesthetic curves aesthetic curves

$$\rho=c_{0}e^{c_{1}s} \tag{2.5}$$

The curve which satisfies Eqn. (2.8) or Eqn. (2.9) is called the log-aesthetic curve.

*2.1.2  Parametric expressions log-aesthetic curves*

In this subsection, we will show parametric expressions of the log-aesthetic curves.

We assume that a curve $C(s)$ satisfies Eqn. (2.8). Then

$$\rho(s) = (c_0 s + c_1)^{\frac{1}{\alpha}} \qquad (2.6)$$

As $S$ is the arc length, $|dC(s)/ds| = 1$ (refer to, for example, [5]) and there exists $\theta(s)$ satisfying the following two equations:

$$\frac{dx}{ds} = \cos\theta, \quad \frac{dy}{ds} = \sin\theta \qquad (2.7)$$

Since $1/(d\theta/ds)$,

$$\frac{d\theta}{ds} = (c_0 s + c_1)^{-\frac{1}{\alpha}} \qquad (2.8)$$

If $\alpha \neq 1$,

$$\theta = \frac{\alpha(c_0 s + c_1)^{\frac{\alpha-1}{\alpha}}}{(\alpha-1)c_0} + c_2 \qquad (2.9)$$

If the start point of the curve is given by $P_0 = C(0)$,

$$C(s) = P_0 + e^{ic_2} \int_0^s e^{i\frac{\alpha(c_0 u + c_1)^{\frac{\alpha-1}{\alpha}}}{(\alpha-1)c_0}} du \, \theta \qquad (2.10)$$

For the second general equation of aesthetic curves expressed by Eqn. (2.9),

$$\frac{d\boldsymbol{\theta}}{ds} = \frac{1}{c_0} e^{-c_1 s} \qquad (2.11)$$

$$\theta = -\frac{1}{c_0 c_1} e^{-c_1 s} + c_2 \qquad (2.12)$$

Therefore the curve is given by

$$C(s) = P_0 + e^{ic_2} \int_0^s e^{-\frac{i}{c_0 c_1} e^{-c_1 u}} du \qquad (2.13)$$

## 2.2  Curvature Based Energy Functionals for Fair Curves and Surfaces

Most surface energy functionals for fair surfaces are related to energy functionals designed for curves. We can design various surface energy functionals by selecting different properties of the curves on the surface to measure curvature or change in curvature and by considering special subsets of the surface curves, i.e. geodesics, lines of curvature [13].

*2.2.1  Bending energy*

The bending energy functional is a generalization of Bernoulli's "elastica" energy that measures the square of curvature $\int \kappa^2 ds$ integrated over the length of a given curve. For curves on a surface, we usually consider only the normal curvature $\kappa_n(\theta)$. This curvature is a function of the principal curvature $(\kappa_{max}, \kappa_{min})$ parameterized by the angle $\theta$ made with the first principal direction. The following functional is often used for the bending energy functional $E_B$ as an area integral of the surface,

$$E_B = \int (\kappa_{max}^2 + \kappa_{min}^2) dA \qquad (2.14)$$

The above equation can be reformulated as follows:

$$E_B = \int (\kappa_{max}^2 + \kappa_{min}^2) = 4 \int H^2 dA - 2 \int G \, dA \qquad (2.15)$$

where $H$ and $K$ are the mean and Gaussian curvatures, respectively. Note that the bending energy is shift-invariant because the area is expanded or shrunk at the square of the dimensional scaling factor, but the square of the curvature is inversely decreased or increased at the same rate.

### 2.2.2    MVS energy

Moreton and Sequin [24] introduced the "MVS" functional that measures curvature variation by integrating the squares of the derivatives of the principal curvatures in the directions of their respective principal directions. The multiplication of the area term is for scale invariance [23].

$$E_{MVS} = \int (\frac{dk_{max}^2}{de_{max}^2} + \frac{dk_{min}^2}{de_{min}^2}) dA \int dA \tag{2.15}$$

where $e_{max}$ and $e_{min}$ are principal curvature directions. at the same rate.

### 2.2.3    $MVS_{cross}$ energy

Joshi and Sequin [13] introduced the "$MVS_{cross}$" functional that adds the change in normal curvature along the in-line direction to the MVS functional.

$$E_{MVS_{cross}} = \int (\frac{dk_{max}^2}{de_{max}^2} + \frac{dk_{min}^2}{de_{min}^2} + \frac{dk_{max}^2}{de_{min}^2} + \frac{dk_{min}^2}{de_{max}^2}) dA \int dA \tag{2.15}$$

We can roughly say that the $MVS_{cross}$ energy of a surface measures the deviation of the surface from a perfect sphere or a cylinder.

As discussed above, many types of the functionals for fair surfaces have been proposed, but for aesthetic design, the designers are usually not using them for practical design because the controllability of the surface deformation is not high enough and it takes a lot of time to minimize the functional for surfaces.

## 2.3    Discrete Filter

For the generation of high-quality surfaces used for car styling design, Farin et al. [6] proposed a surface smoothing method which for each character line of the surface, sequentially selects a point on the curve where the curvature variation criterion introduced by them is the highest in the curve and locally smooth the curve around the point. In their method, first a B-spline curve is fitted to an input sequence of points. Then they extract such a point where the difference of the third derivative, or the derivative of the curvature is the largest and remove a knot corresponding to the point. By repeating this process, they modify the shape of the curve to have a smooth curvature plot whose horizontal axis is the arc length and vertical axis is the curvature. Eck and Jaspert proposed a method to use the difference of the curvature calculated discretely as a local criterion as a fairing method of a sequence of points without B-spline curves [4]. Wagner proposed a method to smooth trajectories of robot manipulation using fourth differences of the sequence of points instead of curvature [25]. Based on the method proposed by Wagner, Higashi and Yamada made it applicable to a curve with a non-uniform knot vector by replacing fourth difference with fourth divided difference and extended it to discrete surfaces which may have defect points [9, 10].

The purpose of the methods mentioned above is for aesthetic design and they can yield curves and surfaces of a certain quality from the view point of the monotonicity of the curvature variation, or smoothness, but they do not remove the curvature instability that exists in polynomial curves like B-spline pointed out by Miura [16]. On the other hand, the log-aesthetic filter [21] can control the curvature. The discrete log-aesthetic filter does not minimize any integral quantities or perturb the positions of the points to minimize any objective functions. It finds locally the most approximate log-aesthetic curve for a given set of points and fits the points to the selected log-aesthetic curve. This is the difference of the methods which have and do not have desired shape targets as the log-aesthetic curve.

To generate a high-quality surface for aesthetic design, it is desirable to solve the true nonlinear minimization problem as Moreton and Sequin [22], and Joshi and Sequin [13] performed. Schneider and Kobbelt [24] solved the nonlinear equation $\Delta H = 0$, and Bobenko and Schroder minimized the discrete Willmore flow [2]. Eigensatz et al. applied a bilateral filter directly on the discrete mean curvature function [5]. At the current state, for surface fairing it is not possible to obtain both of capability of representation flexibility and high quality required for aesthetic design by minimizing objective functions based on variational principle .

In this paper, instead of minimizing a specific functional, we will propose a surface filter which makes the surface have a specified Gaussian curvature distribution in a similar way for the discrete log-aesthetic filter to make the curve have a specified curvature distribution.

## 3    SELF-AFFINITY OF SURFACE

In this and the following section, we think about how to extend the log-aesthetic curve formulation into a surface. The research on the log-aesthetic surface was initiated by Kanaya et al. [14] and Hadara et al. [8] proposed the log-aesthetic curved surfaces, but we cannot say that the formulations of such surfaces have been established.

### 3.1    Correspondences among Differential Geometrical Quantities

Since the log-aesthetic curve describes the relationship between its radius of curvature and arc length, it is necessary to specify some quantities corresponding to them to extend it into a surface. As pointed out by Miura et al. [21], among the arc length $s$, the curvature $\kappa$ and the arc length of its image in the indicatrix of tangents, there is such a relationship that $\kappa = \lim_{s \to 0} \sigma / s$ and it is similar to $K = \lim_{S \to 0} S' / S$ where $K$ is Gaussian curvature, $S$ is the area of a surface $\boldsymbol{S}(u, v)$, and $S'$ is the area of the Gaussian map [4]. Hence we let the curvature of the curve $\kappa$ and the arc length $s$ correspond to the Gaussian curvature $K$ and the surface area $S$, respectively. This implies that the circular arc with a constant curvature corresponds to a surface with a constant Gaussian curvature. By using these correspondences, the fundamental equation of the aesthetic curve $\rho^{\alpha-1} d\rho / ds = C_0$ corresponds to $(1 - K)^{\alpha-1} d(1 / K) / dS = C_1$ where $C_0$ and $C_1$ are arbitrary constants.  Note that

$$\lim_{S \to 0} \frac{S'}{S} = \frac{\iint |N_u \times N_v| du dv}{\iint |S_u \times S_v| du dv} = K \tag{3.1}$$

where $N_u$ and $N_v$ are the derivative of the surface normal with respective to parameters $u$ and $v$, respectively.

The mean curvature $H$ is a surface curvature as important as the Gaussian curvature. For example, the surface with $H = 0$ is called a minimal surface and it is an important example in variational principle [3]. The thin membrane of soap surrounded by an arbitrary boundary is always a minimal surface. However, since the Gaussian curvature $K \le 0$, the surface cannot possess blobby parts and it is frequently inadequate for aesthetic design.  In geometric modeling, for example, as mentioned in Section 1., Schneider and Kobbelt [24] solved the nonlinear equation $\Delta H = 0$. It is necessary to research on filters using the mean curvature in the future.

### 3.2    Self-Affinity of the Plane Curve

We define self-affinity of the plane curve as follows [19]. **Self-affinity of the plane curve**: For a curve generated by removing arbitrary head portion of the original curve, by scaling it with different factors in its tangent and normal directions on every point on the curve, if the original curve is obtained, then the curve has self-affinity. If a given plane curve satisfies $\rho^{\alpha-1} d\rho / ds = C$, the curve has self-similarity of this definition [17, 18].

For a given curve $\boldsymbol{C}(s)$ parameterized by the arc length parameter $s$, we assume the derivative of its curvature, hence that of its radius of curvature as well are continuous. In other words, we assume the curve has $C^3$ continuity. In addition, the radius of curvature $\rho$ is assumed not to be equal to $0$.

By scaling the curve with different factors in the tangent and normal directions (affine transformation of the plane curve [21]), we think about how to make the scaled curve become congruent with the original curve. We therefore re-parameterize the given curve $\boldsymbol{C}(s)$ using a new parameter $t = as + b$ where $a$ and $b$ are positive constants as shown in Fig. 1. To scale the curve uniformly in the tangent direction is equivalent to relate a point $\boldsymbol{C}(t_0 = as_0 + b)$ to another point $\boldsymbol{C}(s_0)$ as shown in Fig.1. In this relationship the scaling factor in the tangent direction $f_t$ is given by $1/a$.

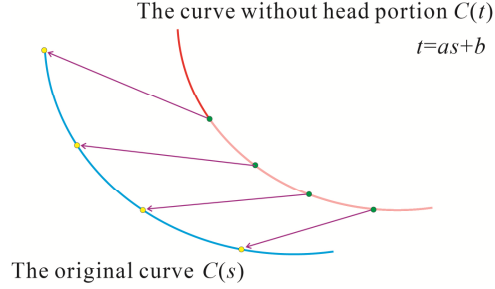

Fig. 1: Self-affinity of the plane curve.

Although $a$ and $b$ are constants, they are related to the scaling factors in the tangent and normal directions $f_t$ and $f_n$ and depend on the shape of the curve. Hence we can not specify them independently. The start point of the curve $C(t)$ is given by $C(b)$ that is a point when $s = 0$. Hence $C(t)$ is a curve without the head portion of the original curve $C(s)$.

The condition can be described for a curve to have self-affinity by the following. **Condition for a plane curve to have self-affinity**: For an arbitrary constant $b > 0$, some $a > 0$ is determined. With these $a$ and $b$, for any $s > 0$ the following equation is satisfied.

$$\frac{\rho(s)}{\rho(as+b)} = f_n \tag{3.2}$$

where $f_n$ is a constant dependent on and determined by $b$ and it is a scaling factor in the normal direction. $f_n$ is given by substituting $s = 0$ in the above equation as follows:

$$f_n = \frac{\rho(0)}{\rho(b)} \tag{3.3}$$

### 3.3 Self-Affinity of the Surface

Although the curve has the arc length parameterization, the surface does not have a suitable parameterization for analysis. However it is possible to have such a local parameterization that the tangent vectors with respect to the parameters are orthogonal each other and their directions are the same as the principal direction and the norm of each tangent vector is equal to 1 [3]. Such a parameterization is called isometric. It is not possible to have a globally isometric parameterization in general. Here we intend to develop a discrete filer for the surface and we assume such a locally isometric parameterization.

We pick up the Gaussian curvature and use its reciprocal of the Gaussian curvature $R = 1/K$ and define the condition for the self-affinity of the surface as follows. **Condition for a surface to have self-affinity**: We assume that the surface is given by $S = S(s,t)$. For an arbitrary constant $b > 0$ and $d > 0$, some $a > 0$ and $c > 0$ are determined. With these $a$, $b$, $c$ and $d$, for any $s > 0$ and $t > 0$, the following equation is satisfied.

$$\frac{R(s,t)}{R(as+b,cs+d)} = f_R \tag{3.4}$$

where $f_R$ is a constant dependent on and determined by only $b$ and $d$, and it does not depend on depend on $s$ or $t$. It is a scaling factor of $R$ and is given by substituting $s = 0$ and $t = 0$ in the above equation as follows:

$$f_R = \frac{R(0,0)}{R(b,d)} \tag{3.5}$$

The scaling factors $f_s$ and $f_t$ in the directions of $S_u$ and $S_v$ are given by $f_s = 1/a$ and $f_t = 1/c$.

If $f_R = 1$, from Eqn. (3.4),

$$R(s,t) = R(as+b,ct+d) \tag{3.6}$$

In this case $R(s,t)$ is constant, i.e. the Gaussian curvature $K$ is also constant. The surface includes a planar, spherical, or cylindrical surface. Furthermore if one of the principal curvature is equal to $0$, it can represent a developable surface including conical and tangent surfaces. The spherical surface is the most typical type of the surfaces of revolution with a positive constant Gaussian curvature. The surface of revolution with $K>0$ includes rugby-ball and barrel types. If $K<0$, it includes the pseudosphere [4].

If $f_R \neq 1$, then Eqn. (3.4) is rewritten as follows:

$$R(s,t) - f_R(b,d)R(as+b, ct+d) = W(s,t) = 0 \qquad (3.7)$$

Hence the function $W(s,t)$ is always equal to $0$. The necessary and sufficient condition for its directional derivative to be always equal to $0$ is $\nabla W(s,t) = 0$. Therefore

$$\frac{\partial W(s,t)}{\partial s} = \frac{\partial R(s,t)}{\partial s} - a\, f_R \frac{\partial R(u, ct+d)}{\partial u}\bigg|_{u=as+b} = \frac{\partial R(s,t)}{\partial s} - a \frac{f_R}{f_s} \frac{\partial R(u, ct+d)}{\partial u}\bigg|_{u=as+b} = 0 \qquad (3.9)$$

$$\frac{\partial W(s,t)}{\partial t} = \frac{\partial R(s,t)}{\partial t} - c\, f_R \frac{\partial R(as+v, v)}{\partial v}\bigg|_{v=ct+d} = \frac{\partial R(s,t)}{\partial s} - \frac{f_R}{f_t} \frac{\partial R(as+d, v)}{\partial v}\bigg|_{v=ct+d} = 0 \qquad (3.10)$$

Eqns. (3.9) and (3.10) are similar to the equation obtained in the curve case [19] and we obtain the following equation:

$$R(s,t) = (c_0 s + c_1)^{\frac{1}{\alpha}} + (c_2 t + c_3)^{\frac{1}{\beta}} \qquad (3.11)$$

where $\alpha = \log f_R / \log f_s$ and $\beta = \log f_R / \log f_t$.

The similar argument for $K$ instead of $R$ works out and the following equation on $K$ can be obtained

$$K(s,t) = (c_0 s + c_1)^{\frac{1}{\alpha}} + (c_2 t + c_3)^{\frac{1}{\beta}} \qquad (3.12)$$

where $\alpha = \log f_K / \log f_s$ and $\beta = \log f_K / \log f_t$ if

$$\frac{K(s,t)}{K(as+b, cs+d)} = f_K \qquad (3.13)$$

## 4    VARIATIONAL FORMULATION

In this section, at first we discuss about the variational principle with a simple example and explain how to formulate the log-aesthetic curve, especially about the functional which the log-aesthetic curve minimizes. Then we extend the functional to formulate the log-aesthetic surface.

### 4.1    Variational Principle

The variational analysis deals with a problem where an objective functional in an integral form should be minimized or maximized. For examples,

$$J = \int_{x_1}^{x_2} f(y, y_x, x)\, ds \qquad (4.1)$$

where $y$ is a function of $x$ and $y_x$ is a derivative of $y$ with respect to x. $y$ is unknown. The condition that $J$ has a stationary value is given by the following partial differential equation:

$$\frac{\partial f}{\partial y} - \frac{d}{dx} \frac{\partial f}{\partial y_x} = 0 \qquad (4.2)$$

This equation is called Euler equation. If $f = f(y, y_x)$, i.e. $f$ is given explicitly without x, the above equation means that

$$f - y_x \frac{\partial f}{\partial y_x} = c \qquad (4.3)$$

where $c$ is a constant.

The simplest example of the variational problem is to minimize the distance between two given points in the x-y plane. An infinitesimal element of the distance is given by

$$ds = \sqrt{(dx)^2 + (dy)^2}\, dx \tag{4.4}$$

and the distance $J$ is given by

$$J = \int_{x_1, y_1}^{x_2, y_2} ds = \int_{x_1}^{x_2} \sqrt{1 + y_x^2}\, dx \tag{4.5}$$

Hence $f(y, y_x, x) = (1 + y_x^2)^{\frac{1}{2}}$ and is given explicitly without x. By Eq.(4.3) we obtain

$$\frac{1}{\sqrt{1 + y_x^2}} = c \tag{4.6}$$

Therefore there exists a constant $a$ such that $y_x = a$. It yields

$$y = ax + b \tag{4.7}$$

where $b$ is a constant as well as $a$. These constants are determined by making the line pass through the given two points $(x_1, y_1)$ and $(x_2, y_2)$.

## 4.2  Variational Formulation of Log-aesthetic Curve

In Eqn. (2.4) if we substitute $\boldsymbol{\rho}^{\boldsymbol{\alpha}}$ with $\boldsymbol{\sigma}$, then the equation is given by

$$\boldsymbol{\sigma} = cs + d \tag{4.8}$$

The above equation means that the log-aesthetic curve is given by a straight line in the $s - \boldsymbol{\sigma}$ plane where the horizontal and vertical axes are the arc length $s$ and $\boldsymbol{\sigma} = \boldsymbol{\rho}^{\boldsymbol{\alpha}}$, respectively to connect two given points $(s_1, \boldsymbol{\beta}_1)$ and $(s_2, \boldsymbol{\beta}_2)$. In this case the following objective functional $J_{LAC}$ is minimized.

$$J_{LAC} = \int_{s_1}^{s_2} \sqrt{1 + \boldsymbol{\sigma}_s^2}\, ds = \int_{s_1}^{s_2} \sqrt{1 + \boldsymbol{\alpha}^2 \boldsymbol{\rho}^{2\boldsymbol{\alpha}-2} \boldsymbol{\rho}_s^2}\, ds \tag{4.9}$$

## 4.3  Variational Formulation of Log-aesthetic Surface

Here we apply the idea of variational principle to the surface formulation. As discussed in Section 3.1, we let the curvature of the curve $\boldsymbol{\kappa}$ and the arc length $s$ correspond to the Gaussian curvature $K$ and the surface area $S$, respectively. In Eqn. (4.9), when $\boldsymbol{\alpha} = -1$, $\boldsymbol{\kappa}_s = -\boldsymbol{\rho}_s / \boldsymbol{\rho}^2$ and we obtain the following equation.

$$J_{LAC} = \int_{s_1}^{s_2} \sqrt{1 + \boldsymbol{\kappa}_s^2}\, ds \tag{4.10}$$

By reparameterizing the above equation with $s = s(t)$, it becomes

$$J_{LAC} = \int_{t_1}^{t_2} \sqrt{x_t^2 + y_t^2 + \boldsymbol{\kappa}_t^2}\, dt = \int_{t_1}^{t_2} \sqrt{\boldsymbol{\lambda}_c^2 + \boldsymbol{\kappa}_t^2}\, dt \tag{4.11}$$

where $s_1 = s(t_1)$, $s_2 = s(t_2)$, and $\boldsymbol{\lambda}_c = \sqrt{x_t^2 + y_t^2}$. Note that $ds/dt = \boldsymbol{\lambda}_c$.

By extending Eqn. (4.11) into the surface, we define the objective functional for the surface $J_{LAC}$ as follows:

$$J_{LAS} = \int_{u_1}^{u_2} \int_{v_1}^{v_2} \sqrt{\boldsymbol{det(I)} + K_u^2 + K_v^2}\, du\, dv \tag{4.12}$$

where $\boldsymbol{I}$ is a matrix expressed with the first fundamental quantities by

$$\boldsymbol{I} = \begin{bmatrix} E & F \\ F & G \end{bmatrix} \tag{4.13}$$

where $E = \boldsymbol{S}_u \cdot \boldsymbol{S}_u$, $F = \boldsymbol{S}_u \cdot \boldsymbol{S}_v$, **and** $G = \boldsymbol{S}_v \cdot \boldsymbol{S}_v$. Note that the area of the surface $S$ is given by

$$S = \int_{t_1}^{t_2} \sqrt{\boldsymbol{det(I)}}\, du\, dv \tag{4.14}$$

As in Section 3.3, we assume a local parameterization $(s, t)$ around a point on the surface $\boldsymbol{S}(s_1, t_1)$ such that the tangent vectors with respect to the parameters are orthogonal each other and their directions are the same as the principal direction and the norm of each tangent vector is equal to 1. With this parameterization, $\boldsymbol{I}$ becomes the $2 \times 2$ unit matrix. By performing integration around the point $\boldsymbol{S}(s_1, t_1)$, Eqn. (4.11) is rewritten as

$$\Delta J_{LAS} = \int_{t_1}^{t_1+\Delta t} \int_{s_1}^{s_1+\Delta t} \sqrt{1+K_s^2+K_t^2} \, ds dt \tag{4.15}$$

According to variational principle, in order to minimize the following functional,

$$J = \int_{t_1}^{t_2} \int_{s_1}^{s_2} g(K, K_s, K_t, s, t) ds dt \tag{4.16}$$

the following equation should be satisfied.

$$\frac{\partial g}{\partial K} - \frac{\partial}{\partial s}\frac{\partial g}{\partial K_s} - \frac{\partial}{\partial t}\frac{\partial g}{\partial K_t} = 0 \tag{4.17}$$

Note that $g = \sqrt{1+K_s^2+K_t^2}$ does not explicitly depend on $K$. Eqn. (4.17) yields

$$(1+K_t^2)K_{ss} - 2K_s K_t K_{st} + (1+K_s^2)K_{tt} = 0 \tag{4.18}$$

The above equation is called the minimal surface or Lagrange's equation and the surface $S(s,t) = (s,t,K(s,t))$ is given by a minimal surface. Therefore in case where the Gaussian curvature on the boundary is specified, the Gaussian curvature should be given by a minimal surface interpolating the boundary values. Please remind that the above discussion assumes a locally isometric parameterization and a globally isometric parameterization does not exist in general. It is not possible to deal with the case where the functional is defined globally as in Eqn. (4.12). In such a case, we should adopt some optimization technique to minimize the functional to generate a target surface.

According to Bernstein's theorem [14], if the boundary of the surface is located infinitely far away, the minimal surface is given by a plane. Therefore the Gaussian curvature is given by

$$K(s,t) = c_0 s + c_1 t + c_2 \tag{4.19}$$

where $c_0$, $c_1$, and $c_2$ are constants. When $\alpha = \beta = 1$ in Eqn. (3.12), the above equation is equivalent to Eqn. (3.12).

For further extension, we may use the mean curvature $H$ instead of the Gaussian curvature $K$ and the similar discussion is also satisfied. In this section we have not discussed about the effects of the powers $\alpha$ and $\beta$. In order to take into account the effects of these powers, we may use $\kappa_{max}^\beta \kappa_{min}^\beta$ where $\kappa_{max}$ and $\kappa_{min}$ are the maximum and minimum normal curvatures, respectively. For example, an objective functional may be defined by

$$J_{LAS} = \int_{u_1}^{u_2} \int_{v_1}^{v_2} \sqrt{det(\boldsymbol{I}) + (\kappa_{max}^\alpha \kappa_{max}^\beta)_u^2 + (\kappa_{max}^\alpha \kappa_{max}^\beta)_v^2} \, du dv \tag{4.20}$$

$$= \int_{u_1}^{u_2} \int_{v_1}^{v_2} \sqrt{det(\boldsymbol{I}) + 2(\kappa_{max}^{2\alpha} \kappa_{min}^{2\beta-1})\{\alpha\kappa_{min}(\kappa_{max,u} + \kappa_{max,v}) + \beta\kappa_{max}(\kappa_{min,u} + \kappa_{min,v})\}} \, du dv$$

These extensions are not dealt in this paper and they are future research topics.

## 5    DISCRETE LOG-AESTHETIC SURFACE FILTER

The discrete log-aesthetic curve filter is constructed based on Eqn. (2.4) [21]. Similarly we construct a discrete surface filter based on Eqn. (3.12) for a triangular-meshed surface. Here we intend to construct an isotropic filter whose effect does not depend on the direction on the tangent plane and we assume $\alpha = \beta$ in Eqn. (3.12).

As the simplest case, we assume $\alpha = \beta = 1$. Note that if $\alpha = \beta = 1$, Eqns. (3.12) and (4.19) are equivalent. Hence the filter has a property to approximate the distribution of the Gaussian curvature by a plane.

We move the location of a vertex $P_i$ in the mesh to satisfy Eqn. (4.19). W restrict the new location $P_i'$ of $P_i$ to satisfy

$$P_i' = P_{ic} + \phi N_i \tag{5.1}$$

where $P_{ic}$ is an average location of the vertices connected to $P_i$ and is a normal vector $N$ there. The coefficients $c_0$, $c_1$, and $c_2$ in Eqn. (4.19) are determined by projecting the positions of the vertices close to $P_i$ to the tangent plane there by use of the values of the Gaussian curvature of these vertices by the least square method. The value of $\phi$ is determined to have the value of the plane at $P_i$.

## 5.1 Curvature Calculation of Mesh

The Gaussian curvature $K$ at the vertex $P_i$ of the triangular mesh is approximately given by

$$K = \frac{a}{S} \tag{5.2}$$

where $a = 2\pi - \sum_{j=0}^{n} \theta_j$ and $S = \sum_{j=0}^{n} S_j / 3$ as shown in Fig. 2. The summation is performed for the triangles around the vertex $P_i$ and $\theta_j$ is an angle of the j-th triangle there and $S_j$ is its area.

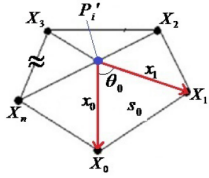

Fig. 2: The definition of $a$ at the vertex $V$.

As shown in Fig. 3, we define $X_i$, i=0,...,n which are the positions of the vertices around the vertex $P_i$ whose new position is $P_i'$ as follows:

$$X_j = (X_{xk}, X_{yk}, X_{zk}) \quad (k = 0, \cdots n) \tag{5.3}$$

$$x_i = X_k - P_i' = X_i - (P_{ic} + \phi N_i) \tag{5.4}$$

The angle between $x_k$ and $x_{k+1}$ is $\theta_k$ and the area of $\triangle P_i' X_k X_{k+1}$ is $s_k$. They are expressed by

$$\theta_k = \cos^{-1}\left(\frac{x_k \cdot x_{k+1}}{|x_k \cdot x_{k+1}|}\right) \tag{5.5}$$

$$s_k = \frac{1}{2}\left\{|x_k|^2|x_{k+1}|^2 - (x_k \cdot x_{k+1})^2\right\}^{\frac{1}{2}} \tag{5.6}$$

where $x_k \cdot x_{k+1}$ denotes the scalar product of the vectors $x_k$ and $x_{k+1}$. If we use $P_{ic}^k = P_{ic} - X_k$, $x_k \cdot x_{k+1}$ is given by a quadratic function of $\phi$ as follows:

$$x_k \cdot x_{k+1} = |N_i|^2 \phi^2 + N_i \cdot (P_{ic}^k + P_{ic}^{k+1})\phi + P_{ic}^k \cdot P_{ic}^{k+1} \tag{5.2}$$

## 5.2 Implementation of the Surface Filter

For the implementation of the surface filter, we use the bisection method to determine in Eqn. (5.1). When the distance between the connected vertices in a given meshed surface is small enough, we can assume that the vertex under processing and its neighborhood vertices are on the same plane. If $\phi$ in Eqn. (5.1) increases, then the Gaussian curvature also increases. Although the Gaussian curvature becomes negative according to the shape of the surface, it is usually possible to determine $\phi$ by extending the search range. In case where a suitable $\phi$ is not found, the vertex $P_i$ is moved to $P_{ic}$.

## 5.3 Application Examples

We measured a plastic car model (1/24 size) and applied our surface filter to a surface of its bonnet part as shown in Fig. 3. We used $\alpha = \beta = 1$ for the surface filter. The number of the vertices in this surface is 8,205 and that of the triangles is 2,735.

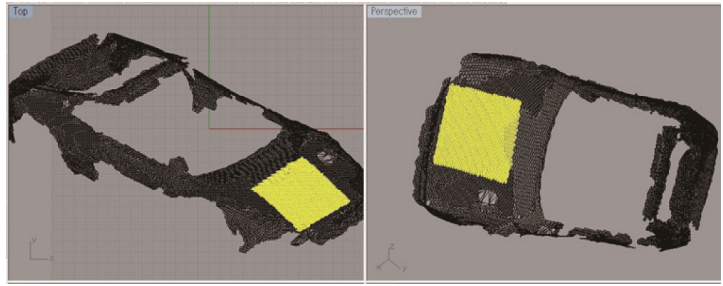

Fig. 3: Measurement data.

Figures from 4 to 7 show the rendering images, the distribution of the Gaussian curvature and zebra mapping before and after filtering. We can see that the noises are removed and the high-quality surface was obtained. We used a PC with Core i7 2.80 GHz and the processing time was 1.59 sec for 10 times filtering.

## 6   CONCLUSIONS AND FUTURE WORK

We have proposed a formulation of the log-aesthetic surface by use of variational principle. We have also implemented a discrete surface filter in the simplest form constructed based on the log-aesthetic surface formulation and we found out that the filter is effective to remove noise and yields high-quality surfaces by applying it to a practical measurement data.

As future work, we will implement optimization codes using Eqn.(4.12) and Eqn.(4.20) to establish the formulation of the log-aesthetic surface.

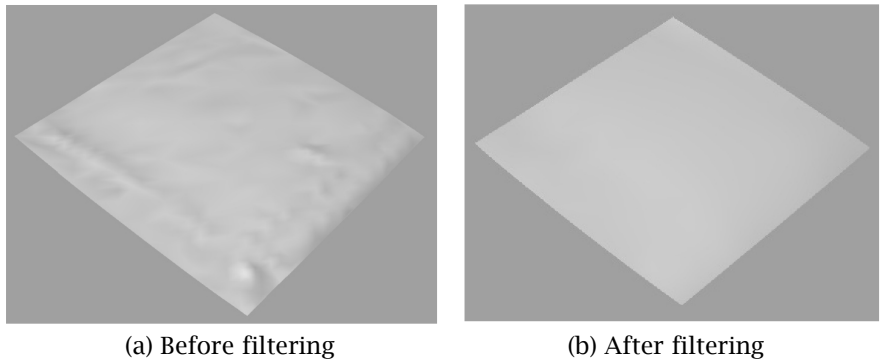

(a) Before filtering          (b) After filtering

Fig. 4: Measurement data.

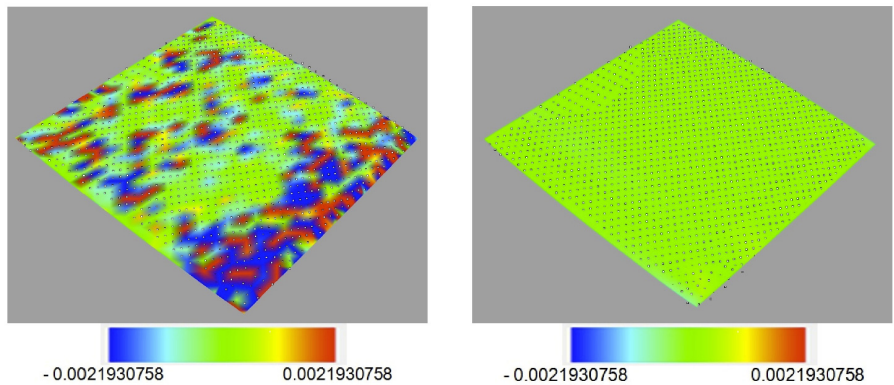

(a) Before filtering          (b) After filtering
Fig. 5: Distribution of the Gaussian curvature

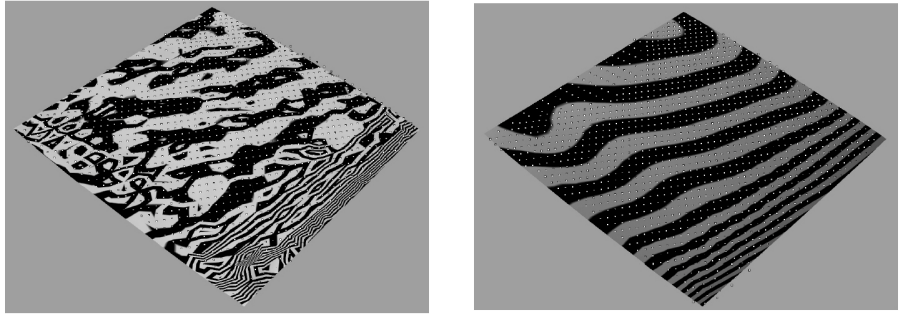

(a) Before filtering          (b) After filtering
Fig. 6: Zebra mapping

ACKNOWLEDGEMENTS

This work was supported in part by grants from the Japan Science and Technology Agency.